\documentclass[PP]{AEA}

\usepackage{natbib}

\newcommand{\ad}{\overset{{\mathrm{a}}}{\sim}}

\newcommand{\Ep}{{\mathrm{E}}}

\draftSpacing{1.5}

\begin{document}

\title{Mastering Panel 'Metrics: Causal Impact of Democracy on Growth}
\shortTitle{Democracy and Growth}
\author{Shuowen Chen and Victor Chernozhukov and Iv\'an Fern\'andez-Val\thanks{Chen: Department of Economics, Boston University, swchen@bu.edu. 
Chernozhukov: Department of 
Economics, MIT, vchern@mit.edu.  Fern\'andez-Val: Department of Economics, Boston University, ivanf@bu.edu. We gratefully acknowledge research support from the British Academy's visiting fellowships, National Science Foundation, and Spanish State Research Agency MDM-2016-0684 under the Mar\'ia de Maeztu Unit of Excellence Program. Fern\'andez-Val was visiting UCL and Cemfi while working on this paper, he is grateful for their hospitality.}}\pubMonth{Month}
\date{\today}
\pubYear{Year}
\pubVolume{Vol}
\pubIssue{Issue}
\maketitle

The relationship between democracy and economic growth is of long standing interest. We revisit the panel data analysis of this relationship by \cite{anrr14} using state of the art econometric methods. We argue that this and lots of other panel data settings in economics are in fact high-dimensional, resulting in principal estimators -- the fixed effects (FE) and Arellano-Bond (AB) estimators -- to be \textit{biased}  to the degree that \textit{invalidates} statistical inference.  We can however remove these biases by using simple analytical and sample-splitting methods, and thereby restore valid statistical inference.  We find that the debiased FE and AB estimators produce substantially higher estimates of the long-run effect of democracy on  growth, providing even stronger support for the key hypothesis in \cite{anrr14}.  Given the ubiquitous nature of panel data, we conclude that the use of debiased panel data estimators should substantially improve the quality of empirical inference in economics.

\section{Mastering Panel 'Metrics}\label{sec:em}
\subsection{The Setting}
 We consider the dynamic linear panel data model 
\begin{equation}\label{eq:LPD}
Y_{it} =a_i + b_t + D_{it}' \alpha + W_{it}'\beta + \epsilon_{it}, 
\end{equation}
where $ i =1,..., N$ and $t =1,...,T$.  Here $Y_{it}$ is the outcome for an observational unit $i$ at time $t$, $D_{it}$ is a vector of variables of interest or treatments, whose predictive effect $\alpha$ we would like to estimate, $W_{it}$ is a vector of covariates or controls including a constant and lags of $Y_{it}$, $a_i$ and $b_t$ are unobserved unit and time effects that can
be correlated to $D_{it}$, and $\epsilon_{it}$ is an error term normalized to have zero mean for each unit and time  that satisfies the weak exogeneity condition
\begin{equation}\label{eq:we}
\epsilon_{it} \perp I_{it}, \ \ I_{it} := \{(D_{is}, W_{is}, b_s)_{s=1}^t, a_i\}.
\end{equation}

%[and is serially uncorrelated conditional on $I_{it}$, the information set of unit $i$ at time $t$.] [CHECK THIS CONDITION AS IT MIGHT RULE OUT CLUSTERING] 

We assume that the vectors 
$
Z_i :=\{(Y_{it}, D_{it}', W_{it}')'\}_{t=1}^T,
$
which collect these variables for the observational unit $i$,  are i.i.d. across $i$, and make other  conventional regularity assumptions.  %This assumption does allow for arbitrary dependence of data for unit $i$ across $t$.  
The main challenge in the estimation of panel data models is how to deal with the unobserved effects.  We review two approaches. 
%In our analysis the temporal dimension $T$ will be small and the cross-sectional dimension $n$ will be large. Accordingly, we shall derive formal asymptotic results under  the ``large $n$, fixed $T"$ asymptotics, were $n \to \infty$ and $T$ is fixed.  This type of scenario is often called the ``short panel".
%The orthogonality condition stated will be strengthened below to various
%assumptions, which permit application of common estimation methods for performing inference on the target parameter $\alpha$.

%The random variable $a_i$ is the unobserved \textit{unit effect}.   It can
%be correlated to $X_{it}$, and so we can not omit it without introducing \textit{omitted variable bias} that leads
%to inconsistent estimates of the parameter of interest $\alpha$.   We can give context to this point
%by thinking of the case where $a_i$ is the unobserved individual's innate ability, $Y_{it}$ is 
% wage, $D_{it}$ is education, and $W_{it}$ are other characteristics of a person $i$ at time $t$.
%Clearly omission of $a_i$ from the model would lead to an omitted variable bias and inconsistent
%estimation of the target parameter $\alpha$ for the usual
%reasons that we discussed in L2.  Figure \ref{fig:ovb} illustrates the omitted variable bias problem in the linear panel model.
%
%Before we continue, it is worth pointing out that  $W_{it}$ could contain a $T$-dimensional vector of indicators
%for time periods as a subvector, namely the vector
%$$Q_{t} = (0,0,...,1,...,0)'$$ with $1$ in the $t$-th position.  In this case the model is said to include \textit{time effects}.

\subsection{The Fixed Effects Approach}
This approach treats the unit and time effects as parameters to be estimated by applying OLS in the model:
$$
Y_{it} = D_{it}'\alpha + X_{it}'\gamma + \epsilon_{it},
$$
where $X_{it} := (W_{it}',Q_i',Q_t')'$, $Q_i$ is an $N$-dimensional vector of indicators for observational units with a 1 in the $i$-th position and 0's otherwise, and $Q_t$ is a $T$-dimensional vector of indicators for time periods with a 1 in the $t$-th position and 0's otherwise. The elements of $\gamma$ appearing in front of $Q_i$ and $Q_t$ are called unit fixed effects and time fixed effects, respectively. The resulting estimator is the fixed effect (FE) estimator.  For our purposes, it can be seen as an exactly identified  GMM estimator with the score function
\begin{equation*}\label{eq:fe}
g(Z_i, \alpha, \gamma) =   \left\{ (Y_{it} - D_{it}'\alpha - X_{it}'\gamma)  M_{it} \right\}_{t=1}^T,
\end{equation*}
where $M_{it} := (D_{it}',X_{it}')'$.

The FE estimator is biased with bias of order $N/(NT) = 1/T$, due to estimation of many ($N$) nuisance parameters with $NT$ observations, and the bias decreases as $T$ becomes large. The estimator approaches the true value $\alpha$ as both $NT$  and $T$ become large, but unfortunately the bias  of the estimator is too big relative to the order $1/\sqrt{NT}$ of the stochastic error, resulting in invalid assessment of statistic significance of the estimates.  This necessitates the use of bias correction to restore the  validity of the statistical inference.

\subsection{The AB Approach}
This approach eliminates the unit effects $a_i$
by taking differences across time and uses moment conditions for the variables in differences.  Specifically, define the differencing operator $\Delta$ acting on doubly indexed random variables $V_{it}$ by  creating 
the difference $\Delta V_{it} =V_{it}-V_{it-1}$.
Apply this operator to both sides of (\ref{eq:LPD}) to obtain:
\begin{equation}\label{eq:DLPD}
\Delta Y_{it} =  \Delta D_{it}'\alpha + \Delta X_{it}'\gamma + \Delta \epsilon_{it},
\end{equation}
where $X_{it} = (W_{it}', Q_t')'$.   Note that by \eqref{eq:we},
\begin{equation*}
\Delta \epsilon_{it} \perp (D_{is},W_{is})_{s=1}^{t-1}, \ \ t = 2,\ldots,T.
\end{equation*}
This means that estimation and inference can be done using an overidentified GMM with score
function
\begin{equation*}\label{eq:ab}
g(Z_i, \alpha, \gamma) = \{(\Delta Y_{it} -  \Delta D_{it}'\alpha - \Delta X_{it}'\gamma) M_{it} \}_{t=2}^T,
\end{equation*}
where $M_{it} = [ (D_{is}',W_{is}')_{s=1}^{t-1}, Q_t']$. 
This is the \cite{arellano:bond} estimator.

The AB  estimator enjoys good properties 
when $T$ is very small,  but when $T$ is even modestly large, it uses
many ($m=O(T^2)$)  moment conditions, which results in a bias of order $m/NT = O(T/N)$,
which can be too large relative to the size of the stochastic error $1/\sqrt{NT}$ of the estimator.
In the latter case statistical inference becomes invalid, and we need to employ bias correction methods to restore its validity.

\subsection{GMM under High Dimensionality and  Need to Bias Corrections}\label{sec:bc}
Both FE and AB are GMM estimators in a high-dimensional regime -- with either the number of nuisance parameters or the number of moment equations being large.

In the FE approach,  the dimension of $\alpha$ is low, but the dimension $p$
of the nuisance parameter $\gamma$ is high. We can approximate this situation as
 $p=\dim(\gamma) \to \infty$ when $n \to \infty$, while $d_{\alpha} := \dim(\alpha)$ is held fixed. In the AB approach, the number of moment conditions, $m = \dim(g (Z_i, \alpha,\gamma)),$
could be high, so we can approximate this situation as $m \to \infty$ when $n \to \infty$.  

In either regime, there exist regularity conditions such that if $(p \vee m)^2$ is small compared to $n$:\footnote{For $a,b \in \Re$, $a \vee b := \max(a,b)$.}
\begin{equation}\label{eq:rate}
(p \vee m)^2/n \to 0 \text{ as  } n \to \infty,
\end{equation}
then the standard approximate normality and consistency results of the GMM estimator continue to hold, namely
\begin{equation}\label{eq:ad}
\sqrt{n}(\hat \alpha - \alpha) \ad N(0, V_{11}),
\end{equation}
where $V_{11}$ is the $d_{\alpha} \times d_{\alpha}$ upper-left block of the asymptotic variance of the GMM estimator corresponding to $\hat \alpha$.\footnote{Sufficient conditions are given, for example, by \cite{nw-09} for GMM problems with $m \to \infty$
and $p$ fixed; and by  \cite{hn-04}, \cite{hk-11} and \cite{fw-17} for nonlinear panel data models where $ m \propto p \to \infty$.} 
%often be improved to requiring
%that the dimension is small compared to the sample size, $p/n \to 0$; see \cite{bcck-15} and \cite{cattaneo-farrell-18}. For exogenous linear models, very strong results 
%were obtained in a sequence of papers by \cite{newey:series,cjn-15}, which also cover the case where $p/n \to c>0$, which gives rise to additional terms in the variance formula.} 

The key rate condition   \eqref{eq:rate} can 
be interpreted as the \textit{small bias condition}.   This condition fails to hold in  the  FE approach where $p^2 = O(N^2 + T^2)$   and    $n=NT$, and in the AB approach when $T$ is large because $m^2 = O(T^4)$   and   $n=NT$.  Both of these failures apply to our empirical setting.
%For example, in the empirical example of Section \ref{sec:emp},  $n= 147$ and $T = 18$.  
%This practical rule may be too rough as a guide in some applications. In such cases we may 
%carry out Monte-Carlo experiments, using data-generating 
%processes that mimic the empirical settings one is facing, to see how
%the methods perform.   

To understand where \eqref{eq:rate} comes from, let us focus on the exactly identified case where $p=m$. An asymptotic second order expansion of $\hat \alpha$ around $\alpha$ gives
$$
\hat \alpha - \alpha =  Z_n/\sqrt{n} +  b/n + r_n,
$$
where $Z_n \ad N(0, V_{11})$, $b = O(p)$ is a first order bias term coming from the quadratic term of the expansion, and $r_n$ is a higher order remainder such as $r_n = O_p( (p/n)^{3/2} + p^{1/2}/n)$. Then, 
\eqref{eq:ad} holds
if both
$$
\sqrt{n} b/n \to 0,  \quad \text{i.e. } p^2/n \to 0,
$$
and
$
\sqrt{n}r_n \to_P 0$, i.e.  $p^{3/2}/n \to 0.$

The sketch above illustrates that the bias is the bottleneck. If we remove
the bias somehow, then we can improve the rate requirement
 \eqref{eq:rate} to a weaker condition listed below.  

There are several ways
of removing the bias:
\begin{itemize}
\item[a)] \textit{Analytical bias correction}, where we estimate $b/n$ using analytical expressions for the bias and set $$\check \alpha = \hat \alpha - \hat{b}/n.$$
\item[b)] \textit{Split-sample bias correction}, where we split the sample into two parts, compute the estimator on the two parts
$\hat \alpha_{(1)}$ and $\hat \alpha_{(2)}$ to obtain $\bar \alpha = (\hat \alpha_{(1)}
+  \hat \alpha_{(2)})/2$, and then set
$$
\check \alpha = \hat \alpha - (\bar \alpha - \hat \alpha) = 2 \hat \alpha - \bar \alpha.
$$
In some cases we can average over many splits to reduce variability.\footnote{
In some cases it is also possible to use the bootstrap and leave-one-out methods for bias correction.}
\end{itemize}
Why does the sample-splitting method work?  Assuming that we estimate the same number of nuisance parameters and use the same number of moment conditions in all the parts of the sample, and that these parts are homogenous, then the first order biases
of $\hat \alpha$,  $\hat \alpha_{(1)}$, and $\hat \alpha_{(2)}$ are 
$$
\frac{b}{n}, \quad \frac{b}{n/2}, \quad \frac{b}{n/2},
$$
so that the first order bias  of $\check \alpha$ is 
$$
2 \frac{b}{n} - \left (\frac{1}{2}\left[ \frac{b}{n/2} \right]+ \frac{1}{2} \left[\frac{b}{n/2}\right] \right ) = 0.
$$

After debiasing, the resulting rate conditions are weaker. In particular, there exist regularity conditions such that if  the dimensionality is not overly high:
$$(p \vee m)^{3/2}/n \to 0 \text{ as  } n \to \infty,$$
then the approximate normality and consistency results 
for the bias-corrected GMM estimator continue to hold:\footnote{Sufficient conditions are given in \cite{kiviet95}, \cite{kuerst:febias} and \cite{cpy18} for dynamic linear panel data models and \cite{fw-16} and \cite{fw-17} for nonlinear panel data models.}
$$
\sqrt{n}(\check \alpha - \alpha) \ad N(0, V_{11}).
$$

\subsection{The Debiased FE and AB Estimators}

To construct analytical debiased FE estimator (DFE-A), we need to characterize the first order bias. An analysis similar to \cite{nickell} yields that first order bias $b/n$ obeys:
$$
H b = - \frac{1}{T} \sum_{i=1}^N \sum_{t=1}^{T-1} \sum_{s=t+1}^{T} \Ep[D_{is} \epsilon_{it}], 
$$
for
$$
 H =    \frac{1}{NT} \sum_{i=1}^N \sum_{t=1}^T  \Ep[\widetilde{D}_{it} \widetilde{D}_{it}'],
$$
where $\widetilde{D}_{it}$ is the residual of the sample linear projection of $D_{it}$ on $X_{it}$.
 Note that  $b = O(N)$ because the source of the bias is the estimation of the $N$ unit fixed effects and the order of the bias is $b/n = O(T^{-1})$ because there are only $T$ observations that are informative about each unit fixed effect.\footnote{There is no bias coming from the estimation of the time fixed effects because the model is linear and we assume independence across $i$.}  An estimator of the bias can be formed as
$$
\hat H \hat b = - \sum_{i=1}^N  \sum_{t=1}^{T-1} \sum_{s=t+1}^{(t+M) \wedge T} \frac{ D_{is} \hat \epsilon_{it} }{ T-s+t}, 
$$
where $\hat \epsilon_{it}$ is the fixed effect residual, 
$$
\hat H =    \frac{1}{NT} \sum_{i=1}^N \sum_{t=1}^T  \widetilde{D}_{it} \widetilde{D}_{it}',
$$
and $M$ is a trimming parameter such that $M/T \to 0$ and $M\to\infty$ as $T \to \infty$ \citep{hk-11}.
 
 To implement debiasing by sample splitting, we need to determine the partition of the data. For the debiased FE estimator via sample splitting (DFE-SS), we split the panel along the time series dimension because the source of the bias is the estimation of the unit fixed effects. Thus, following  \cite{dj15}, the parts contain the observations  $\{i = 1,\ldots,N; t = 1, \ldots, \lceil T/2 \rceil \}$ and $\{i = 1,\ldots,N; t =  \lfloor T/2 \rfloor, \ldots, T \}$, where $\lceil \cdot \rceil $ and $\lfloor \cdot \rfloor$ are the ceiling and floor functions. This partition preserves the time series structure and delivers two panels with the same number of unit fixed effects, where there are $T/2$ observations that are informative about each unit fixed effect.   For the debiased AB estimator via sample splitting (DAB-SS), we split the panel along the cross section dimension because the source of the bias is the number of moment conditions relative to the sample size. Thus, the parts contain the observations  $\{i = 1,\ldots, \lceil N/2 \rceil ; t = 1, \ldots, T \}$ and $\{i =  \lfloor N/2 \rfloor,\ldots,N; t =  1,\ldots,T \}$. This partition delivers two panels where the number of observations relative to the number of moment conditions is half of the original panel. Note that  there are multiple possible partitions because the ordering of the observations along the cross section dimension is arbitrary. We can therefore average across multiple splits to reduce variability. 
  
% . In this case since there is no bias coming from the estimation of the time effects, $D=0$, the jackknife bias-corrected estimator simplifies to
%\begin{equation}\label{eq:dynamic}
%\check \alpha = \hat \alpha - (\hat \alpha_{N, T/2} - \hat \alpha) = 2\hat \alpha - \hat \alpha_{N, T/2}.
%\end{equation}

\section{Democracy and Growth}\label{sec:emp}
%\subsection{Data}
We revisit the application to the causal effect of democracy on economic growth of \cite{anrr14} using the econometric methods described in Section \ref{sec:em}. To keep
the analysis simple, we use a balanced sub-panel of 147 countries over the period from 1987 through 2009 extracted from the data set used in  \cite{anrr14}.  The outcome variable $Y_{it}$ is the logarithm of GDP per capita in 2000 USD as measured by the World Bank for country $i$ at year $t$. The treatment variable of interest $D_{it}$ is a democracy indicator constructed in \cite{anrr14}, which combines information from several sources including Freedom House and Polity IV.  It characterizes whether countries have free and competitive elections, checks on executive power, and an inclusive political process. We report some descriptive statistics of the variables used in the analysis in the online supplemental Appendix.  

%% latex table generated in R 3.2.4 by xtable 1.8-2 package
%% Sun Apr 10 16:20:44 2016
%\begin{table}[ht]
%\centering
%\caption{Descriptive Statistics }\label{table:ds_democracy}
%\begin{tabular}{lcccc}
%  \hline\hline
% & Mean & SD & Dem = 1 & Dem = 0 \\ 
%  \hline
%Democracy & 0.62 & 0.49 & 1.00 & 0.00 \medskip \\ 
%  Log(GDP) & 7.58 & 1.61 & 8.09 & 6.75 \medskip \\  \hline
%  Number Obs. & 3,381 & 3,381 & 2,099 & 1,282 \\ 
%   \hline\hline
%\end{tabular}
%\end{table}

We control for unobserved country effects, time effects and rich dynamics of GDP using the linear panel model \eqref{eq:LPD}, where $W_{it}$ includes four lags of $Y_{it}$. The weak exogeneity  
condition \eqref{eq:we} implies that democracy and past GDP are orthogonal to contemporaneous and future GDP shocks $\epsilon_{it}$ and that these shocks
are  serially uncorrelated (since $W_{it}$ includes the lagged values of $Y_{it}$).   

In addition to the instantaneous or short-run effect of a transition to democracy to economic growth measured by the coefficient $\alpha$, we are interested in a permanent or long-run dynamic effect.  This effect in the dynamic linear panel model \eqref{eq:LPD}  is 
\begin{equation}\label{eq:lr}
\alpha/(1 - \sum_{j=1}^4 \beta_j),
\end{equation}
where $\beta_1, \ldots, \beta_4$ are the coefficients corresponding to the lags of $Y_{it}.$

We consider the FE and a one-step AB estimators as well as their debiased versions (DFE and DAB).      Indeed, the raw AB and FE fail to satisfy the small bias condition: the AB approach relies on $m=632$ moment conditions  to estimate  $p = 169$ parameters with  $n=147 \times 18 = 2,646$  observations, after using the first five periods as initial conditions, so that $(m \vee p)^2/n \approx  150$, which is not close to zero; the FE  approach estimates $p=170$ parameters with  $n=147 \times 19 = 2,793$  observations, after using the first four  periods as initial conditions, yielding $(m \vee p)^2/n \approx  10$, 
which is not close to zero.

To debias the estimators, we  consider both analytical and split sample bias corrections. For the fixed effect approach, DFE-A  implements the analytical debiasing with $M=4$, whereas DFE-SS implements debiasing by sample splitting.  We consider two versions of the debiasing via sample-splitting for AB, where DAB-SS1 uses one random split and DAB-SS5 uses the average of five random splits. 

For each estimator, we report analytical standard errors clustered at the country level and bootstrap standard errors based on resampling countries with replacement.   The estimates of the long-run effect are obtained by plugging-in estimates of the coefficients in the expression \eqref{eq:lr}. We use the delta method to construct  analytical standard errors clustered at the country level, and resample countries with replacement to construct bootstrap standard errors. There is no need to recompute the analytical standard errors
for debiased estimators, because the ones obtained for the uncorrected estimators remain valid for the bias corrections. We also report bootstrap standard errors for the debiased estimators.

%:
Table \ref{table:democracy} presents the empirical results.\footnote{We obtained the estimates with the commands \texttt{plm} and \texttt{pgmm} of the package \texttt{plm} in \texttt{R}.}  
 FE   finds that a transition to democracy increases economic growth by almost 1.9\% in the first year and 16\% in the long run, while  AB  finds larger impacts of 4\% and 21\% but less precisely estimated. 
We find that the debiasing changes the estimates by a significant amount in 
both statistical sense (relative to the standard error) and economic sense (relative to the uncorrected estimates). The debiased  estimators, DFE and DAB, find that a transition to democracy increases economic growth by about 2.3-5.2\% in the first year, and about 25-26\% in the long run.  Interestingly, the two debiased approaches produce very similar estimates.  Moreover, the results coincide with the results obtained using the method of \cite{hhk05}, as reported in \cite{anrr14}.  We believe that the estimates reported here as well as the later estimates reported in \cite{anrr14} represent an adequate, state of the art analysis. Of course, it would be interesting to continue to explore other modern, perhaps even more refined, econometric approaches to thoroughly examine the empirical question.

We conclude with comments on the standard errors. The analytical standard errors are smaller than the bootstrap standard errors for the split-sample bias corrections. These  differences might indicate that the analytical standard errors miss the additional  sampling error introduced by the estimation in smaller panels.The analytical correction produces more precise estimates than the split-sample correction. 

\begin{table}[ht]
\centering
\caption{Effect of Democracy on Economic Growth}\label{table:democracy}
\begin{tabular}{lcccccc}
  \hline\hline
    \multicolumn{1}{l}{} & \multicolumn{3}{c}{Initial and Debiased FE} & \multicolumn{3}{c}{Initial and Debiased AB } \\
  & FE & DFE-A  & DFE-SS &  AB & DAB-SS1 & DAB-SS5 \\ 
  \hline
Short Run Effect 
& 1.89 & 2.27 &  2.44 & 3.94 & 5.22 & 4.53 \\ 
  of Democracy  & (0.65)  &  &  & (1.50) &  &  \\ 
    \ \ \ \   ($\times 100$) & [0.64] &  [0.64]  & [0.96] & [1.52] & [1.83] & [1.91]  \medskip \\ 
  1st lag of log GDP & 1.15 & 1.23 &  1.30 & 1.00 & 0.98 & 1.03 \\ 
   & (0.05)  & &    & (0.06) &  &  \\ 
   & [0.05] & [0.05] & [0.08] & [0.06] & [0.07] & [0.08] \medskip \\ 
 2nd lag of log GDP  & -0.12 & -0.14 & -0.13 & -0.06 & -0.05 & -0.07 \\ 
   & (0.06)   & &  & (0.06) &  &  \\ 
   & [0.05] & [0.05]   & [0.08] & [0.06] & [0.07] & [0.07] \medskip \\ 
 3rd lag of log GDP  & -0.07 & -0.09  &  -0.13 & -0.04 & -0.04 & -0.06 \\ 
   & (0.04)   & &   & (0.04) &  &  \\ 
   & [0.04] & [0.04]  & [0.06]  &  [0.04] & [0.04] & [0.04] \medskip \\ 
 4th lag of log GDP  &  -0.08 & -0.08 &  -0.08 & -0.08 & -0.08 & -0.08 \\ 
   & (0.02)   & &  & (0.03) &  &  \\ 
   & [0.02] &  [0.03]  & [0.04] & [0.03] & [0.03] & [0.03] \medskip \\ \hline
   Long-run effect  & 16.05 & 25.91  & 25.69 &  20.97 & 26.46 & 25.24  \\ 
   of democracy & (6.67) &   & & (9.51) &  &  \\ 
    \ \ \ \   ($\times 100$) & [6.63] & [9.31] & [12.12] & [9.38] & [10.72] & [11.29] \\ 
  \hline\hline
%   \multicolumn{8}{l}{\footnotesize{Note 1: All the specifications include country and year effects.}}\\
%   \multicolumn{8}{l}{\footnotesize{Note 2: Clustered standard errors at the country level in parentheses.}}\\
%   \multicolumn{8}{l}{\footnotesize{Note 3: Bootstrap standard errors in brackets based on 500 replications.}}\\
 \end{tabular}
 \begin{tablenotes}
All the specifications include country and year effects. Analytical clustered standard errors at the country level are shown in parentheses. Bootstrap standard errors based on 500 replications
are shown in brackets.
\end{tablenotes}

\end{table}

\bibliographystyle{aea}
\bibliography{mybibVOLUME}

%\appendix

\section*{Appendix}
The online supplemental Appendix contains  the data, descriptive statistics,  and code in \texttt{R} and \texttt{Stata} for the empirical application. 

\appendix

\section{Supplemental Appendix}

% latex table generated in R 3.2.4 by xtable 1.8-2 package
% Sun Apr 10 16:20:44 2016
\begin{table}[ht]
\centering
\caption{Descriptive Statistics }\label{table:ds_democracy}
\begin{tabular}{lcccc}
  \hline\hline
 & Mean & SD & Dem = 1 & Dem = 0 \\ 
  \hline
Democracy & 0.62 & 0.49 & 1.00 & 0.00 \medskip \\ 
  Log(GDP) & 7.58 & 1.61 & 8.09 & 6.75 \medskip \\  \hline
  Number Obs. & 3,381 & 3,381 & 2,099 & 1,282 \\ 
   \hline\hline
\end{tabular}
\end{table}

\end{document}